\documentclass[draft]{agujournal2019}
\usepackage{url} 
\usepackage{lineno}
\usepackage[inline]{trackchanges} 
\usepackage{soul}
\usepackage{amsmath}
\draftfalse

\journalname{JGR: Space Physics}

\begin{document}

%
%


\title{Short wavelength electrostatic wave measurement using MMS spacecraft}

\authors{Ahmad Lalti\affil{1,2}, Yuri V. Khotyaintsev\affil{1}, Daniel B. Graham\affil{1}}

\affiliation{1}{Swedish Institute of Space Physics, Uppsala, Sweden}
\affiliation{2}{Space and Plasma Physics, Department of Physics and Astronomy, Uppsala University, Uppsala, Sweden}

\correspondingauthor{Ahmad Lalti}{ahmadl@irfu.se}




\begin{keypoints}
\item We show that electric field measurements of short-wavelength waves aboard MMS are distorted.
\item We develop a method to get reliable measurements of the 3D wave vector of such waves.
\item Application of this method to solar wind ion acoustic waves shows them to be predominantly field-aligned with wavelength around 10 Debye. lengths
\end{keypoints}

%
%

%
%


\begin{abstract}
Determination of the wave mode of short-wavelength electrostatic waves along with their generation mechanism requires reliable measurement of the wave electric field. We investigate the reliability of the electric field measurement for short-wavelength waves observed by MMS. We develop a method, based on spin-plane interferometry, to reliably determine the full 3D wave vector of the observed waves. We test the method on synthetic data and then apply it to ion acoustic wave bursts measured in situ in the solar wind. By studying the statistical properties of ion acoustic waves in the solar wind we retrieve the known results that the wave propagation is predominantly field-aligned. We also determine the wavelength of the waves. We find that the distribution peaks at around 100 m, which when normalized to the Debye length corresponds to scales between 10 and 20 Debye lengths.
\end{abstract}

\section{Introduction}

    Most plasma environments in space are collisionless \cite{baumjohann2012basic}, which means that interparticle collisions do not play any significant role in the dynamical evolution of the system. In such a case, and if we ignore the effect of gravitational forces, long range electromagnetic forces govern the dynamics of the plasma. One of the still remaining open questions in space plasma physics is the question of irreversible energy dissipation without collisions. It is believed that short wavelength electrostatic waves, through wave-particle interactions, play a key role in creating such irreversible dissipation whether at collisionless shocks \cite{sagdeev1966cooperative,wilson2014quantified1,wilson2014quantified2}, at magnetic reconnection \cite{khotyaintsev2019collisionless} or at terminating the energy cascade in plasma turbulence \cite{2008Valentinicrossscale,2009Valentinielectrostatic,valentini2014nonlinear}. The exact channels with which such waves dissipate energy is still under investigation.
    
    The study of short wavelength electrostatic waves in space plasmas requires reliable measurement of the wave electric field. Several techniques have been developed for measuring it \cite{mozer1973analyses}. Arguably the most successful of which is the double probe technique \cite{1967Fahleson,pedersen1998electric}, where the electric field is estimated by taking the difference between the probe to spacecraft potential measured at two points in space, then dividing by the probe to probe separation:
    \begin{equation}
        E_{ij} = - \frac{V_i - V_j}{d_{ij}},
        \label{eq1}
    \end{equation}
    where the indices refer to the measurement points (i.e. the probes), $E_{ij}$ is the electric field pointing from probe $j$ to probe $i$ and $d_{ij}$ is the separation distance between probes $i$ and $j$. The first usage of the double probe technique was in 1967 on a sounding rocket flown to the auroral ionosphere \cite{mozer1967electric}. Following this success the double probe technique has been and is currently being used on multiple spacecraft throughout the heliosphere such as the Van Allen probes \cite{wygant2013electric}, THEMIS \cite{bonnell2009electric}, Cluster \cite{gustafsson1997electric}, and Solar Orbiter \cite{maksimovic2020solar} to name a few.
    
    One of the most recent and most advanced missions is the Magnetospheric MultiScale (MMS) mission \cite{MMS_overview}. MMS is a constellation of four spacecraft in a tetrahedral formation equipped with high temporal and spatial resolution particles and fields instruments allowing scientists to probe kinetic scale space plasma phenomena. The Electric field double-probe (EDP) instrument aboard MMS consists of two orthogonal spin-plane double probes (SDP) \cite{lindqvist2016spin} with probe-to-probe distance of 120 m, and axial double probes \cite{ergun2016axial} with probe-to-probe distance of 28.15 m. The set of 6 probes enables the measurement of the 3-dimensional electric field from DC up to 128 kHz. Since its launch in 2015 many studies have used the EDP data from MMS to study plasma wave phenomena in the various plasma regions around Earth from bow shock \cite{2018Vasko,2018Goodrich,wang2021electrostatic,2022Vasko} to the magnetopause \cite{2019Steinvall,2020khotyaintsev,2022Graham} to the magnetotail \cite{2017Lecontel,2021Richard}.
    
    Despite this success, measurement of wave electric field with wavelength comparable to the probe-to-probe distance ($\sim10^2 m$) can become unreliable. That is why there is a need to test the performance of the EDP instrument when it comes to measuring such waves, and understand what effects can affect the electric field measurement. Then develop a method that mitigate those problems to give reliable measurement of the plasma wave properties. In section \ref{measurement} of this paper, we use electrostatic waves whose properties are generally known, namely ion acoustic waves in the solar wind, to test the performance of the EDP instrument on MMS. In section \ref{method} we develop a method to mitigate the problems identified in section \ref{measurement} and reliably measure the full 3D wave vector of the observed waves. In section \ref{test}, we use this method to conduct a case study of one ion acoustic wave burst with peculiar power spectral density (PSD) signature. We also study statistically the properties of those waves. Finally, in section \ref{conclusions} we summarise and conclude.

\section{Short wavelength electric field measurement by MMS}
\label{measurement}

In order to test the performance of MMS when it comes to measuring short-wavelength electrostatic waves, we need to use waves whose characteristics and properties, namely angle of propagation with the background magnetic field, $\theta_{kB}$, wavelength and frequency, are well known. One such wave mode is the ion acoustic waves in the solar wind \cite{gurnett1977plasma,1978Gurnettionacoustic,mozer2020large,pivsa2021first}. Many studies have shown that those waves propagate predominantly in a field aligned direction; e.g. recently \citeA{pivsa2021first} used the Solar Orbiter spacecraft to study ion acoustic waves in the solar wind and found that 80\% of their observed waves have a $\theta_{kB}$ less than or equal to 20 degrees. Furthermore, due to their short wavelength, ion acoustic waves in the fast flowing solar wind can be highly Doppler shifted, so their observed frequency in the spacecraft frame can range from a few hundred to thousands of Hz. 

We compile a set of 210 ion acoustic wave bursts observed with MMS in the quiet undisturbed solar wind. When compiling the list of wavebursts we make sure that the spacecraft was not in the ion or electron foreshock by inspecting the power spectral density (PSD) data product, which resolves frequencies reaching 100 kHz but sampled at lower temporal resolutions. We also make sure that we don't have any Langmuir waves present in the time interval. This way we make sure that the most likely observed wave mode is indeed ion acoustic waves \cite{1978Gurnettionacoustic}.

\begin{figure*}[ht]
    \centering
    \includegraphics[scale=0.35]{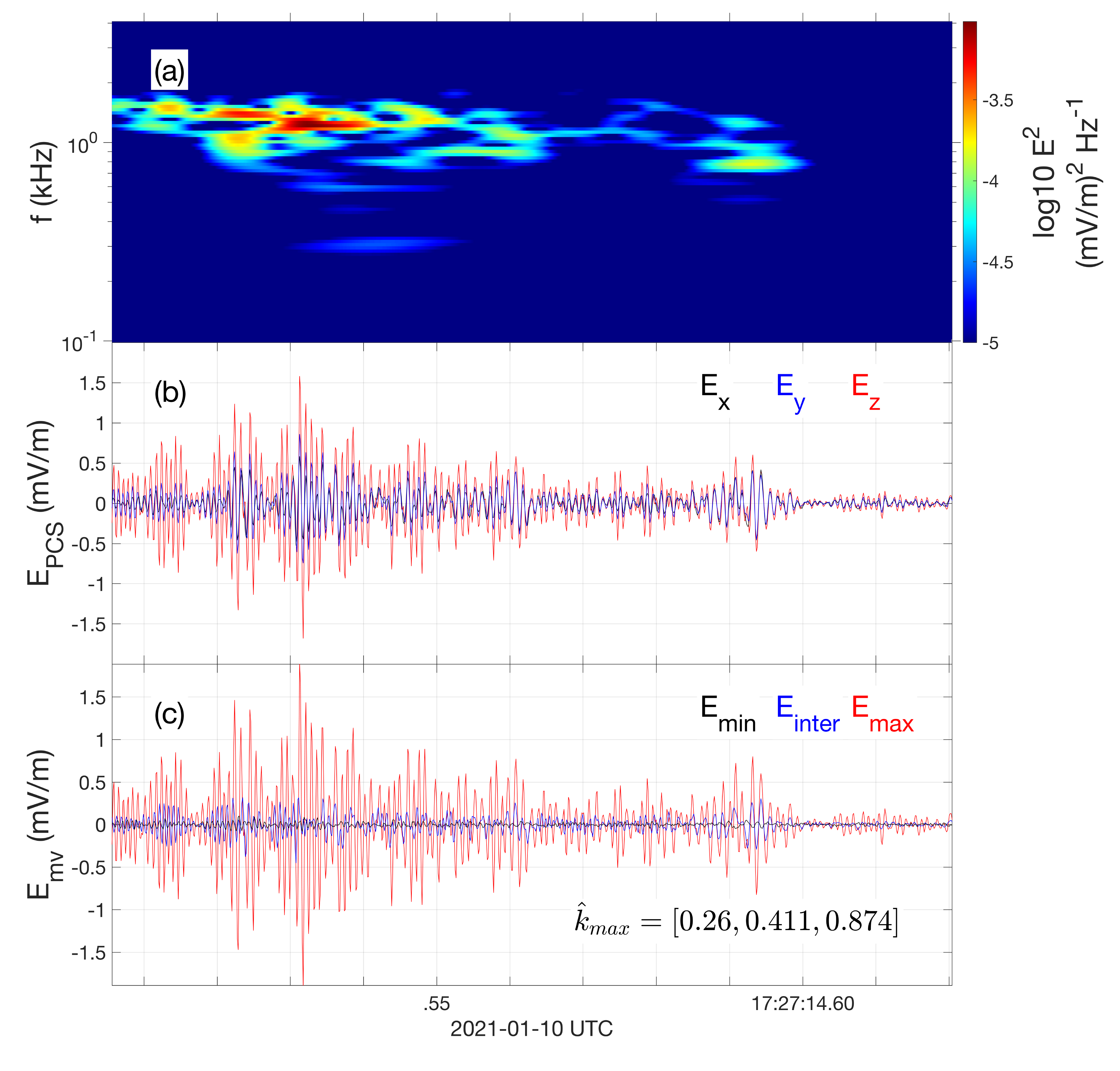}
    \caption{Example of an ion acoustic waveburst observed by MMS1 in the solar wind. Panel (a) shows the PSD of the electric field, (b) shows the electric field measured using the double probe technique and set in the PCS, (c) shows the measured electric field rotated into the coordinate system defined by the minimum, intermediate and maximum variance directions. Overlayed on panel (c) is the maximum variance direction in the PCS. }
    \label{fig:fig1}
\end{figure*}

An example of an ion-acoustic wave burst observed by MMS1 is shown in Figure \ref{fig:fig1}. Panel (a) shows the power spectral density (PSD) of the electric field showing the peak frequency around 1.5 kHz. Panel (b) shows the electric field as measured using the double probe technique (equation \ref{eq1}), and expressed in the probes coordinate system (PCS), shown in Figure \ref{fig:fig3} (a), where the positive x is along probe 1, positive y is along probe 3, positive z is along probe 5 and the spacecraft is located at the origin. Panel (c) of Figure 1 shows the electric field rotated into the coordinate system defined by the minimum, intermediate and maximum variance directions. The maximum variance direction of this waveburst is $\mathbf{\hat{k}}_{max} = [0.26,\;0.41,\;0.87]$, with a large component in the z direction. With a background magnetic field direction of $\mathbf{\hat{b}} = [0.67,\;0.52,\;0.53]$ we obtain $\theta_{kB} = 31^\circ$, which is relatively oblique compared to what is statistically expected in the solar wind.  Having no wave activity near the plasma line (not shown), the only wavemode that fits the characteristics for this event is that of oblique ion acoustic waves. 

If the electric field measurement were accurate one could trust the above results for the mode determination, but three effects can distort the electric field measurement using the double probe technique. The first is the sheath impedance effects \cite{gurnett1998principles,hartley2016using}. The probes are not directly coupled to the plasma, a sheath forming around the probe, causes a potential drop between the plasma and the surface of the probe. In a circuit diagram this sheath can be modeled as a capacitor and resistor connected in parallel between the plasma and the probes \cite<see for example Figure 5 in >[]{hartley2016using}. This parallel RC circuit has a voltage divider effect on the probe with complex impedance, $V_{in}/V_{out} = \frac{Z_L}{Z_s+Z_L}$ where $V_{in}$ is the plasma potential, $V_{out}$ the measured potential and $Z_{L}$/$Z_{s}$ are the load (spacecraft)/sheath impedances, respectively. At low frequencies the probe is resistively coupled to the plasma, and since by design the load resistance is much larger than any expected sheath resistance, the gain defined as the ratio $V_{in}/V_{out}$ is close to 1. On the other hand, at higher frequencies the probe is coupled capacitively to the plasma. In that limit the gain is different from 1 so the measured electric field will exhibit both amplitude attenuation and phase shift \cite{hartley2016using}.

The second effect is the boom shorting effect \cite{pedersen1998electric,califf2016}. Both axial and spin plane probes are connected to a preamplifier, which in turn is connected to the spacecraft by a long conducting wire boom. This boom is grounded to the spacecraft, so when an external electric field exist, it will induce a charge distribution on its surface to satisfy the constant potential boundary condition. This will short out the external electric field causing a decrease in the amplitude of the measured electric field \cite{califf2016}. This decrease in amplitude is due to $d_{ij}$ in equation \ref{eq1} deviating from the physical probe to probe separation, and an effective length rather than geometric length is required when calculating the electric field so the amplitude is not attenuated \cite{pedersen1998electric}.

\begin{figure*}[ht]
    \centering
    \includegraphics[scale=0.35]{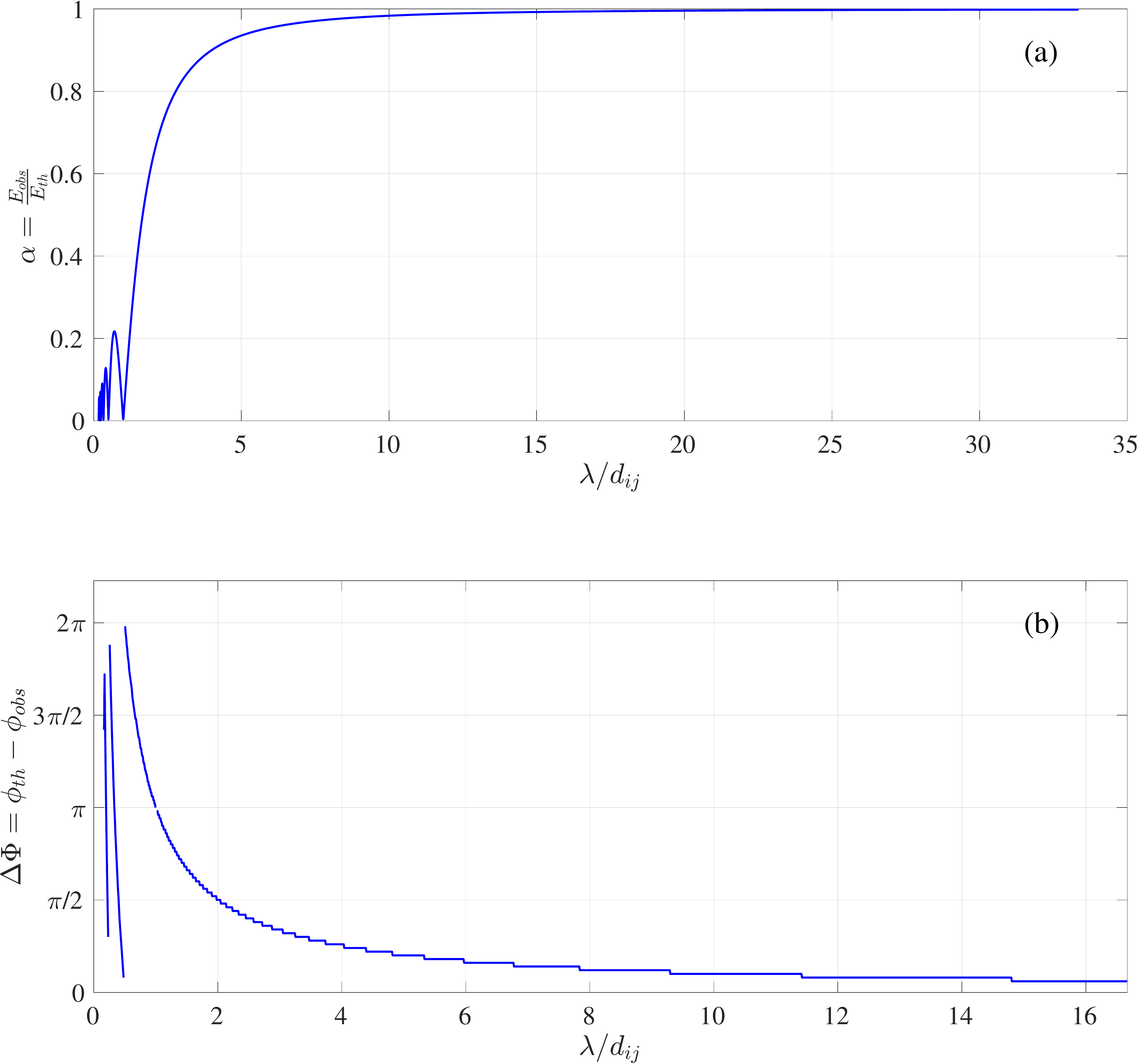}
    \caption{Simulating the short wavelength effect on electric field measurement. (a) shows the attenuation factor $\alpha = E_{obs}/E_{th}$ and (b) the phase shift $\Delta \Phi = \phi_{th} - \phi_{obs}$ versus the normalized wavelength $\lambda/d_{ij}$.}
    \label{fig:fig2}
\end{figure*}

\begin{figure*}[ht]
    \centering
    \includegraphics[scale=0.35]{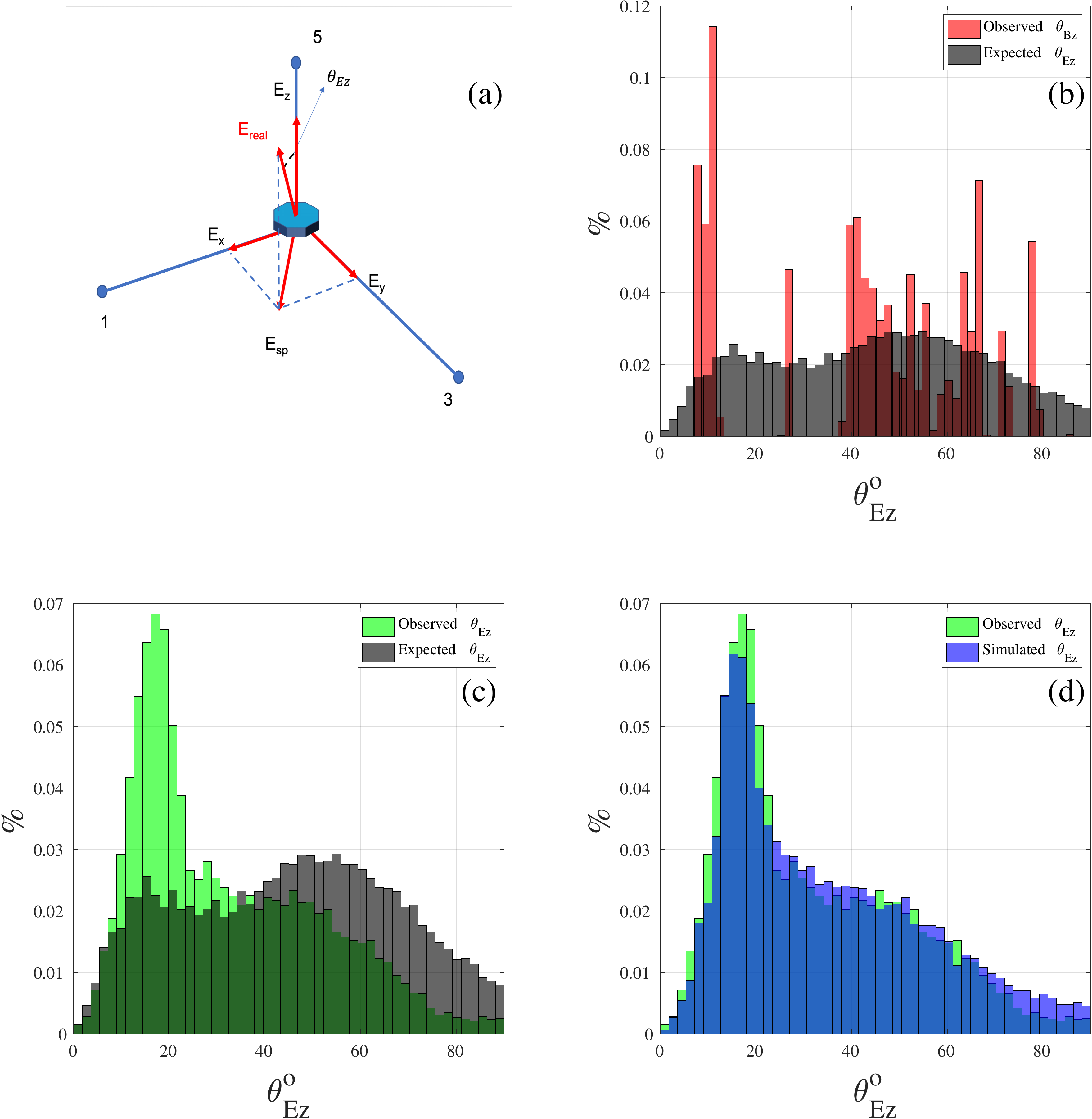}
    \caption{Systematic bias of the electric field measurement towards the axial probes direction. (a) shows schematics of the probes coordinate system. (b) shows the distribution of the angle that the magnetic field makes with the axial direction $\theta_{Bz}$ in red and the expected $\theta_{Ez}$ distribution in black. (c) shows the observed, in green, and the expected, in black,  $\theta_{Ez}$ distributions. (d) shows the observed in green, and the simulated, in blue,  $\theta_{Ez}$ distributions.}
    \label{fig:fig3}
\end{figure*}

Finally, the third effect is the short wavelength effects \cite{labelle1989measurement,gurnett1998principles}. The double probe technique works fairly well at approximating the electric field in cases where the wavelength of the waves $\lambda$ is significantly larger than the probe to probe separation $d_{ij}$. As $\lambda$ approches $d_{ij}$ the electric field starts to be attenuated in amplitude and is phase shifted. To see how this effect works, we simulate a plane wave traveling in the direction of two probes separated by a distance $d_{ij}$ and we vary the wavelength of the wave (Figure \ref{fig:fig2}). We calculate the ratio of the amplitudes of the observed ($E_{obs}$) to the theoretical ($E_{th}$) electric fields $\alpha = E_{obs}/E_{th}$,  from here on we refer to $\alpha$ as the attenuation factor (panel a). We also calculate the phase shift between the observed and theoretical electric fields $\Delta \Phi = \phi_{th} - \phi_{obs}$ (panel b). It is clear from panel (a) that at long wavelengths (large $\lambda/d_{ij}$), $\alpha$ is approximately 1, and as the wavelength approaches the probe to probe separation ($\lambda/d_{ij} \rightarrow 1$), $\alpha$ decreases until it reaches zero at $\lambda/d_{ij} = 1$, then oscillates between positive values and zero for smaller wavelengths. Similar behavior is seen in the phase shift shown in panel (b) where the phase difference increases until it reaches $\pi$ when $\lambda/d_{ij} = 1$.

All three effects cause the underestimation of the observed electric field amplitude. The first two effects, when present, affect the spin-plane probes in the same way, since they were built symmetrically. But since the axial probes have different length and geometry compared to the spin plane probes, the magnitude of the attenuation of the electric field is expected to be different between the spin plane and the axial probes. As for the third effect, since it's highly dependent on the wavelength of the wave in the direction of the probes, the attenuation will be different in all three directions.

By exploiting this asymmetry between the axial and spin-plane probes we can use the angle that the vector electric field $\mathbf{E}$ makes with the z direction, $\theta_{Ez}$, as a measure of the instrument performance. In analyzing the wave bursts we use the probes coordinate system (PCS). For every vector measurement for the 210 wave bursts we calculate $\theta_{Ez}$ we get a total of around 16000 measurement. We compare the observed distribution of $\theta_{Ez}$ to that expected. The former can be obtained directly from the measurement, while to get the latter we calculate the angle between the magnetic field and the axial direction, $\theta_{Bz}$. By adding a spread to the observed direction of the background magnetic field for each wave burst in our list in such a way to simulate the expected distribution of the $\theta_{kB}$ in the solar wind \cite{pivsa2021first}, we get the expected distribution of $\theta_{Ez}$ for the observed waves. 
The red histogram in panel (b) of figure \ref{fig:fig3} shows the distribution of $\theta_{Bz}$, while the expected $\theta_{Ez}$ distribution is shown in black in the same panel. In panel (c) we compare the observed distribution (green) to the expected one (black), and we clearly see that the electric field of short-wavelength waves measured by MMS is systematically shifted towards the axial direction. This is highly problematic as this indicates that the magnitude and direction of the measured electric field are not reliable. So, $\theta_{kB}$ determined for the event shown in Figure \ref{fig:fig1} cannot be trusted.

In order to get a sense of which of the three effects is the dominant we simulate the spacecraft measurement of the observed electric field by launching plane waves with potential profile $V = V_0 \cos (\omega t - \mathbf{k} \cdot \mathbf{x})$, where $V_0$ is the amplitude of the wave, $\omega$ its angular frequency, $\mathbf{k}$ is the wave vector, and $\mathbf{x}$ the spatial location in the PCS. We then measure the potential at the location of the 6 probes, and use it to calculate the electric field in the three probe directions. We can then compare the observed distribution of $\theta_{Ez}$ to the simulated distribution which accounts only for the short wavelength effect. The simulation requires knowledge of the wave vector of the observed waves. Although at this stage we don't have this information, we can use a reasonable approximation. For the direction of propagation, as before (Figure \ref{fig:fig3} (b-c)), we take the direction of the measured magnetic field and add a spread in such a way that the distribution of $\theta_{kB}$ conform to what was observed in \citeA{pivsa2021first}. As for the wavelength, we assume a linear dispersion relation of the form
\begin{equation}
    \omega_{sc} = \mathbf{V}_{sw} \cdot \mathbf{k},
    \label{dr}
\end{equation}
where $\omega_{sc}$ is the measured frequency in the spacecraft frame, $\mathbf{V}_{sw}$ is the solar wind velocity, $\mathbf{k}$ is the wave vector that we want to determine. In using this formula we assume that the waves are highly Doppler shifted by the solar wind bulk motion, which is a reasonable assumption in the solar wind if we ignore waves with $\mathbf{k}$ almost perpendicular to the solar wind velocity \cite{gurnett1991waves}. In this way we get an approximate range of the wave vector $\mathbf{k}$ for each of the wave bursts in our list. 

Figure \ref{fig:fig3} (d) shows the resulting simulated distribution of $\theta_{Ez}$ (blue) overlayed on top of the observed distribution (green). We observe an excellent agreement between the observed and simulated distributions even without accounting for the two other effects. This indicates that the short-wavelength effect has the greatest effect in distorting the measured electric field of short wavelength waves.

\section{3D wave vector determination using spin plane interferometry}
\label{method}

In order to determine the wave mode of any waveburst, it is necessary to determine, in addition to its frequency, its direction of propagation and its wavelength. One of the easiest ways to get the direction of propagation of an electrostatic wave is by determining its maximum variance direction \cite{sonnerup1988minimum}, as was done for the waveburst shown in Figure \ref{fig:fig1}, which coincides with the direction of propagation of electrostatic waves (albeit with $\pi$ ambiguity). But with the systematic shift of the electric field towards the axial direction shown in the previous section, the maximum variance direction of the electric field will not coincide with the direction of propagation of the wave, so there is a need for a reliable method to determine the wave properties despite the technical limitations described earlier.

In this section we develop a method, based on the spin plane multi-probe interferometry, to obtain the full 3D wave vector of any measured electrostatic short wavelength waveburst. 
\begin{enumerate}
    \item We first calculate the frequency dependent wave vector in the spin plane using spin-plane interferometry.
    \item We then use this result to correct for the attenuation effect in the spin plane components of the electric field.
    \item We calculate the maximum variance direction of the corrected electric field and use it to determine the frequency dependent 3D wave vector.
    \item By simulating a plane wave at different propagation directions and different wavelengths and then applying the steps described above to measure the simulated wave properties, we obtain a look-up table that allows us to determine the actual wavelength and propagation direction of a wave from its measured properties.
\end{enumerate}
   
In the following subsections we provide a detailed description of each step of the method.

\subsection{Spin plane wave vector determination}

One powerful method that can be used to determine the wave vector is single spacecraft interferometry. This method has been used before in analyzing short scale waves from various spacecraft and throughout the heliosphere \cite<>[to cite a few]{1996Bonnell,2004vaivads,2005Balikhin,2010khotyaintsev,2016Graham}. The method works by measuring the same quantity (electric field, probe potential, density etc.) at two different locations in space. When a localized structure or a plane wave passes the spacecraft, it will leave a signature in the measured quantity at one location and then at the other depending on its direction of propagation. By measuring the time delay $\Delta t$ (or equivalently the phase shift $\Delta \phi$) between the two measurements and knowing the distance between the two measurement points $\mathbf{d}$, one can determine the wave vector ($\mathbf{k}$) in the direction of the two measurement points using:
\begin{equation}
    \mathbf{k} \cdot \mathbf{d} = \Delta \phi.
    \label{phase}
\end{equation}

\begin{figure*}[ht]
    \centering
    \includegraphics[scale=0.65]{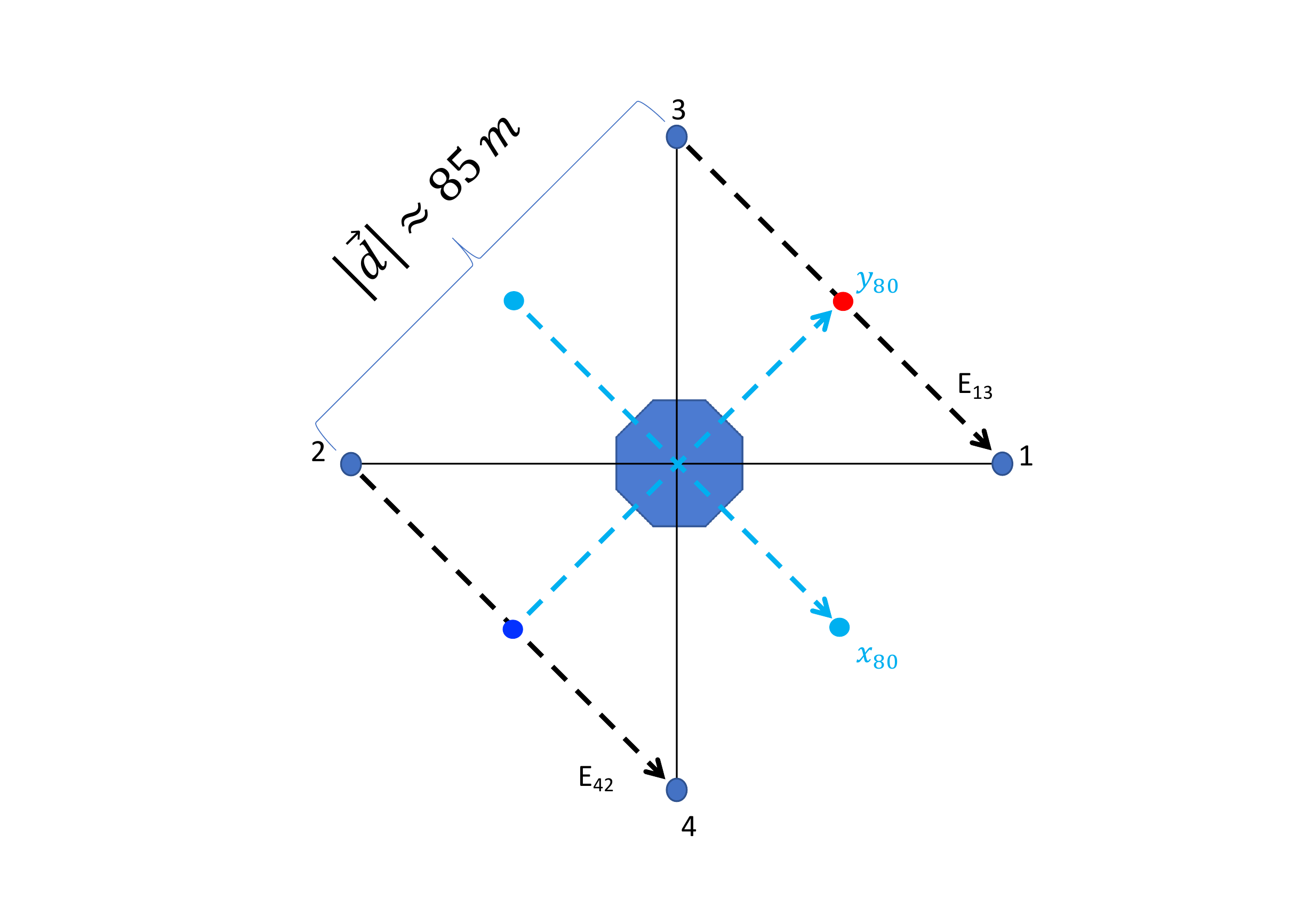}
    \caption{Schematics of the spin plane probes showing the two orthogonal directions along which interferometry is applied, forming the E80 coordinate system.}
    \label{fig:fig4}
\end{figure*}

For electrostatic waves measured by MMS, one can apply interferometry on both the probe potentials or electric fields. Using synthetic data, \citeA{steinvall2022applicability} found that one particular electric field configuration, what they call ``diagonal electric field'' and what we will call E80 electric field, is the most reliable quantity to apply interferometry on. In order to explain the E80 interferometry we show in Figure \ref{fig:fig4} a schematic of MMS in the spin plane, using probes 2 and 4 we calculate the electric field $E_{42}$, and using probes 3 and 1 we calculate the electric field $E_{13}$. Those two electric fields are the same field components measured at two different spatial locations (the dark blue and red circles in the schematics) separated by a distance $d_{80} \approx 85$ m. So by measuring the phase shift between the two measurements one can use equation \ref{phase} to obtain the wave vector component in the direction named $y_{80}$ in Figure \ref{fig:fig4}. The same thing can be applied to the orthogonal direction where phase shifts between electric fields $E_{32}$ and $E_{14}$ can be used to obtain the component of the wave vector in the direction named $x_{80}$ in Figure \ref{fig:fig4}. Using this method one obtains the spin-plane wave vector components. It is worth noting that those two directions ($x_{80}$ and $y_{80}$) define a coordinate system (which we will call E80 coordinate system) that is rotated by $45^\circ$ clock-wise from the coordinate system aligned with the two wire boom pairs (PCS). 

Unfortunately, the E80 interferometry technique is restricted to the spin plane, so we cannot use it to get the axial component of the wave vector. Other quantities can be timed in the axial direction, such as the potential measured at probes 5 and 6, but such timing is often unreliable \cite{steinvall2022applicability}.

Before going in to the 3D wave vector determination, it is informative to see the spin plane interferometry technique in action. And before applying to real data we use synthetic data. We generate a wave packet comprised of a sum of sinusoidal waves all traveling in the same direction at a polar angle $\theta = 60^\circ$ and an azimuthal angle $\phi = 30^\circ$ in the PCS. The waves have frequencies that are logarithmically spaced between 100 Hz and 3.5 kHz, and follow the dispersion relation $f = V_{ph} k/(2\pi) = V_{ph}/\lambda$, where $V_{ph}$ is a constant representing the phase velocity of the waves. Throughout this paper we fix $V_{ph}$ to be equal to $90$ km/s giving a range of wavelength $\lambda \in [25 \; 900] $ m and wavenumber $k \in [0.007\;  0.2477]$ m$^{-1}$.
The potential of the full wave packet has the functional form:
\begin{equation}
    V\left(\mathbf{r},t\right) = \sum_i V_0\sin \left(\omega_i t - \mathbf{k}_i\cdot \mathbf{r}\right),
    \label{eq4}
\end{equation}
where the sum is over all frequency components, $\mathbf{k}$ is the wave vector, $\mathbf{r} = \left(x,y,z\right)$ is the position vector and $V_0$ is a constant amplitude.

\begin{figure*}[ht]
    \centering
    \includegraphics[scale=0.35]{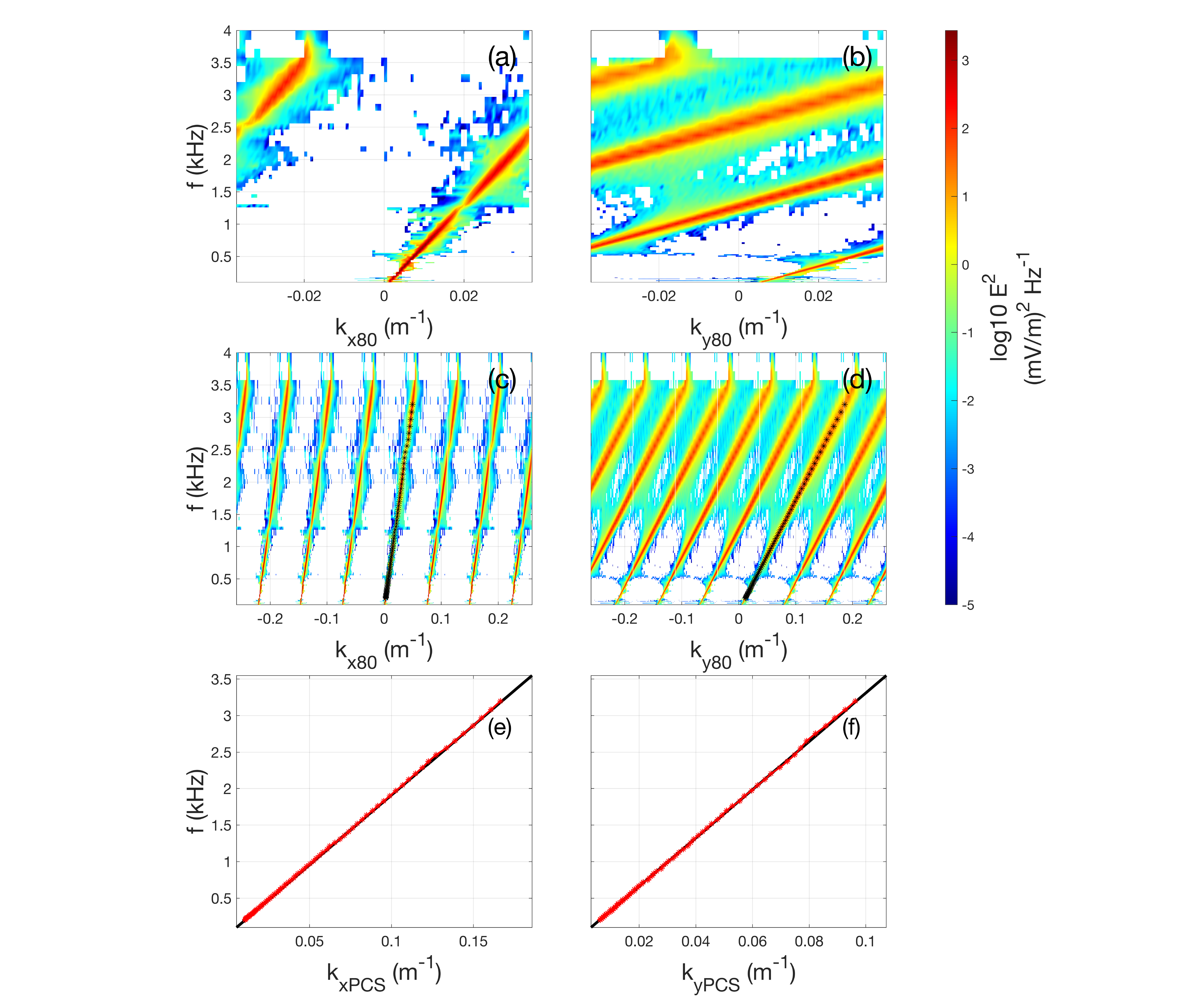}
    \caption{Results of spin plane interferometry. Panels (a) and (b) shows the f - k PSD in the $x_{80}$ and $y_{80}$ directions without accounting for aliasing. Panels (c) and (d) repeats what is plotted in (a-b) but after accounting for aliasing. Panels (e) and (f) shows the theoretical (in black) and measured (in red) dispersion relations in the $x_{PCS}$ and $y_{PCS}$ directions.}
    \label{fig:fig5}
\end{figure*}

We evaluate the potential of this wave packet at 6 different spatial locations that coincide with the location of probes 1 through 6. 
In order to apply interferometry for all the frequency components in the wave packet, we apply a wavelet transform on the electric fields ($E_{42}$ and $E_{13}$ to get the wave vector in the $y_{80}$ direction and $E_{32}$ and $E_{14}$ to get the wave vector in the $x_{80}$ direction). In the wavelet space we can calculate the phase difference between the two signals, $\Delta \phi$ \cite<equation 2 in >[]{2016Graham}, and calculate a wavenumber at each frequency using equation \ref{phase} and each time step. Then we bin the power in frequency - wavenumber (f - k) space by summing the power of every measurement point that has f - k values within each bin. 
The resulting f - k power spectrum for both E80 directions is shown in Figure \ref{fig:fig5} panels (a - b). Those panels show a clear linear dispersion relation marked out with the high values of the PSD.

Interferometry works well for wavelengths larger than twice the distance between the two measurement points, in our case $\lambda > 2 \times d_{80} = 2 \times 85 = 170$ m. Or in terms of the wavenumber $k_{x,y} = 0.037$ m$^{-1}$ where $k_{x,y}$ is the wavenumber in the x/y direction. When the wavelength becomes less than 170 m the measurement will be subject to spatial aliasing. In f - k space aliasing causes the signal to wrap around to the opposite limit of the domain as is seen in panels (a - b) of Figure \ref{fig:fig5}, when the wavenumber approaches the limiting value of 0.037 the dispersion relation continues from the -0.037 value and the same behavior repeats with every encounter of the dispersion relation with the edge of the $k_{x,y} = [-0.037 \; 0.037]$ m$^{-1}$ domain. This effect can easily be mitigated by extending the $k_{x,y}$ domain as is done in panels (c - d), we see how the continuous linear dispersion relation is retrieved. In both panels we see multiple repeating branches. The branch that has a zero intercept (the branch that connects to k = 0) is the one that corresponds to the dispersion relation of the physical wave (equation \ref{dr}).

After having manually selected the appropriate branch and in order to retrieve a single dispersion relation at each frequency we select values of k that correspond to the maximum power. The selected values are shown as black stars in panels (c - d) in Figure \ref{fig:fig5}. Since the main coordinate system that we are using is the PCS, we rotate the spin plane wave vector measured from the E80 to the probe coordinate system. In panel (e) of Figure \ref{fig:fig5} we plot the expected dispersion relation projected along the x direction in the PCS in black. We overlay in red the prescribed dispersion relation obtained by the above described interferometry technique. We do the same in panel (b) but for the y direction. It is clear that after accounting for spatial aliasing we can retrieve the spin plane dispersion relation of an observed wave packet down to wavelengths shorter than the length at which aliasing is expected to occur (170 m). It is worth noting that the dips in power seen in the PSD of Figure \ref{fig:fig5} (especially in panels a and c) are due to frequencies where the projected wavelength along the probes direction is equal to the probe to probe separation of 120 m where we expect the attenuation factor $\alpha$ to be zero.

The conclusion from this subsection is that after accounting for the spatial aliasing in the way described one can reliably retrieve the spin-plane dispersion relation. This method works under one condition only, namely, the wave packet under investigation is sufficiently dispersive to see a dispersion relation in the f - k PSD in order to be able to select the correct branch that connects to the origin in f - k space.

\subsection{3D wave vector determination: Simulation}

The E80 interferometry described in the previous subsection is a reliable method to determine the spin plane components of the wave vector $\left(\mathbf{\kappa}(f_i) = \left[k_x(f_i), \;k_y(f_i)\right]\right)$. To calculate the component of $\mathbf{k}$ along the axial direction we develop the method detailed below:
\begin{enumerate}
    \item Correct for the attenuation effect in the spin plane components:
\begin{itemize}

    \item We calculate the electric field in the PCS ($\mathbf{E}$) such that: $E_x = - \frac{V_1 - V_2}{120}$, $E_y = - \frac{V_3 - V_4}{120}$ and $E_z = - \frac{V_5 - V_6}{28.15}$.
    
    \item Knowing the wavelength at each frequency $f_i$ from the E80 interferometry, we determine the electric field amplitude attenuation factor $\alpha(f_i)$ by interpolating the values from the numerical relations shown in Figure \ref{fig:fig2}a.
    
    \item We Fourier transform the spin-plane components of $\mathbf{E}$ and we correct the amplitude by dividing it by the corresponding $\alpha(f_i)$. We then inverse Fourier transform to obtain a corrected electric field $\mathbf{E}_c$ (the spin-plane components corrected).
    
\end{itemize}
    \item Determine the direction of propagation of the wave at each frequency while accounting for the phase shift due to the attenuation effect:
\begin{itemize}
    \item For each $f_i$ we band pass filter $\mathbf{E}_c$ in a frequency interval $[0.99 \; 1.01]\times f_i$. We calculate the phase shift $\Delta \Phi$ at this frequency by interpolating the values from the numerical relations shown in Figure \ref{fig:fig2}b. If a component has $\Delta \Phi > \pi$ the sign of this component is incorrect, and we correct it by flipping its sign. 
    
    \item To the filtered signal obtained above, we apply a minimum variance analysis in order to obtain the maximum variance direction $\mathbf{M}$, which corresponds to the direction of propagation for electrostatic waves.
     As $\mathbf{M}$ has a $\pi$ ambiguity, we choose its sign such that its spin-plane component is aligned with $\mathbf{\kappa}$.
\end{itemize}

\item Calculate the 3D wave vector ($\mathbf{k}$) at each frequency:
\begin{itemize}
    
    \item We calculate the angle between $\mathbf{M}$ and the axial direction $\theta_{kz}$ and assume that it corresponds to the angle that $\mathbf{k}$ makes with the axial direction.
    
    \item We construct the 3D wave vector by using $\mathbf{\kappa}$ for the spin plane components and determining the axial component using:
    \begin{equation}
         k_z = (k_x^2+k_y^2)^{(1/2)}\frac{\cos(\theta_{kz}) }{\sin(\theta_{kz})}. 
    \end{equation}
    So the final wave vector becomes $\mathbf{k}(f_i) = \left[\mathbf{\kappa}(f_i), \; k_z(f_i)\right]$.
\end{itemize}    
\end{enumerate}

\begin{figure*}[ht]
    \centering
    \includegraphics[scale=0.3]{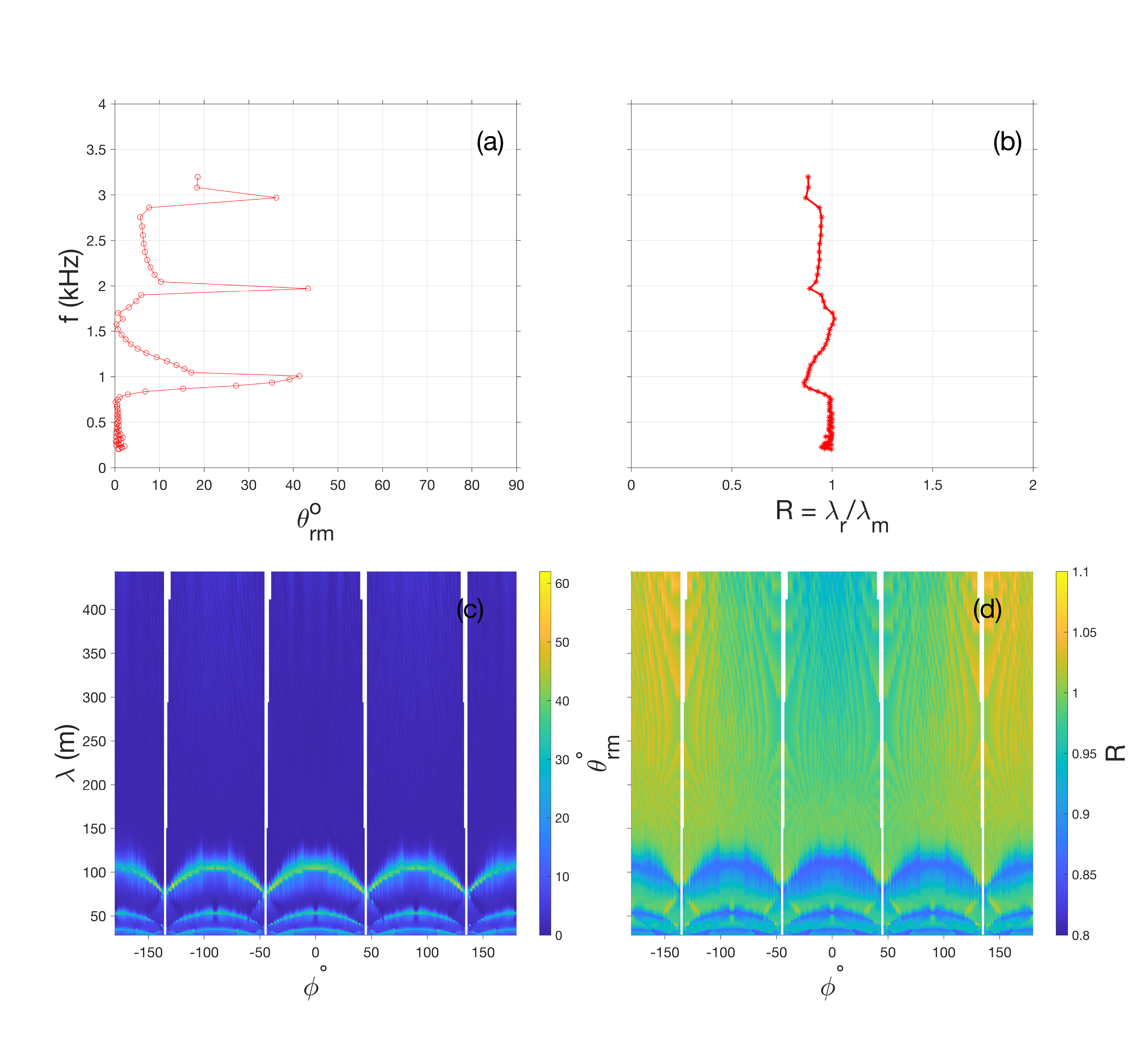}
    \caption{Comparing measured and real properties of the waves. Panel (a) and (b) show the angle between the measured and the real wave vectors $\theta_{rm}$ and the ratio between the real and measured wavelength $R = \lambda_r/\lambda_m$, respectively, for the wave modeled in Figure \ref{fig:fig5}. Panels(c) and (d) show a surface plot of the wavelength $\lambda$ versus the azimuthal angle $\phi$ for waves propagating at a polar angle $\theta_z = 60^\circ$, with the colorbars representing $\theta_{rm}$ and $R$ respectively.}
    \label{fig:fig6}
\end{figure*}

In following those steps we assume that the electric field in the axial direction is not subject to short wavelength attenuation or phase shift.
To check how well the method works, and to what extent those assumptions hold, we plot in Figure \ref{fig:fig6} (a) the angle between the measured and the expected wave vectors $\theta_{rm}$ and in panel (b) the ratio between the expected and measured wavelength $R = \lambda_r/\lambda_m$ of the same waveburst analyzed in Figure \ref{fig:fig5}. For small frequencies (large wavelengths) ($<800$ Hz) $\theta_{rm} <2^\circ$ and $R \approx 1$ showing that for this frequency range the method works well in measuring the 3D wave vector (both direction and magnitude) where both assumptions made are valid. As the wavelength approaches the probe to probe separation (and its integer fraction, $\lambda = d_{ij}/n$ where n is an integer) the method slightly overestimates the wavelength with $R \sim 0.9$ , while greatly deviates when it comes to the direction of propagation with $\theta_{rm} \sim 50^\circ$.

 To check what range in parameter space (all combinations of the triplet $\left(\lambda, \theta, \phi\right)$ ) we expect the method to work we run the simulation above by varying the polar angle $\theta$ of the wave packet in the range $[0 \; 180]^\circ$ and the azimuthal angle $\phi$ in the range $[-180\;180]^\circ$ with an angular resolution of $1^\circ$. For each combination of the triplet $\left(\lambda, \theta, \phi\right)$ we calculate $R$ and $\theta_{rm}$. For a fixed polar angle of $\theta = 60^\circ$, we plot $\theta_{rm}$ in panel (c) and $R$ in panel(d). We see that for all wavelengths larger than 120 m the method works fine in measuring the full 3D wave vector where $\theta_{rm} <2^\circ$ and $R \sim 1$. For wavelengths shorter than 120 m the measured direction of propagation deviates from the real direction of propagation with $\theta_{rm}$ reaching $60^\circ$ while the wavelength is only slightly overestimated with $R$ having a minimum value of 0.8. This occurs only at combinations of $\left(\lambda, \theta, \phi\right)$ where the projection of $\lambda$ on the direction of one of the probes is close to the probe-to-probe separation and its integer fraction \cite{labelle1989measurement}. At such values $\alpha$ in that direction would be close to zero, and the method fails since we are dividing by it to correct for the electric field.

\subsection{3D wave vector determination: The inverse problem}

The results in the previous subsection are informative but when the method is applied to real data one cannot know \textit{a priori} if the waves are in a parameter range where the method is expected to work or not. We use the simulation data to develop a look-up table which we use to determine the 3D wave vector of any measured electrostatic waveburst. We know that we have 5 independent observables: 
\begin{enumerate}
    \item The measured wavelength of the wave $\lambda_{m}$,
    \item The polar angle of the measured wave vector $\theta_{km}$,
    \item The azimuthal angle of the measured wave vector $\phi_{km}$,
    \item The polar angle of the uncorrected electric field $\theta_{Eu}$,
    \item The azimuthal angle of the uncorrected electric field $\phi_{Eu}$.
\end{enumerate}

We then ask the question: what simulated plane wave characterized by the triplet $\left(\lambda, \theta, \phi\right)$ would give us values of the 5 observables that are closest to what we measure from the real wave. We then search for solutions that satisfy the following conditions:
\begin{equation}
\begin{split}
     |\Delta \lambda_{m}| = |\lambda_{m-dat} - \lambda_{m-sim}| \le 5 m,\\
     |\Delta \theta_{m}| = |\theta_{km-dat} - \theta_{km-sim}| \le 5^\circ,\\
     |\Delta \phi_{m}| = |\phi_{km-dat} - \phi_{km-sim}| \le 5^\circ,\\
    |\Delta \theta_{Eu}| = |\theta_{Eu-dat} - \theta_{Eu-sim}| \le 10^\circ,\\
    |\Delta \phi_{Eu}| = |\phi_{Eu-dat} - \phi_{Eu-sim}| \le 10^\circ.
    \end{split}
    \label{condition}
\end{equation}

\begin{figure*}[ht]
    \centering
    \includegraphics[scale=0.35]{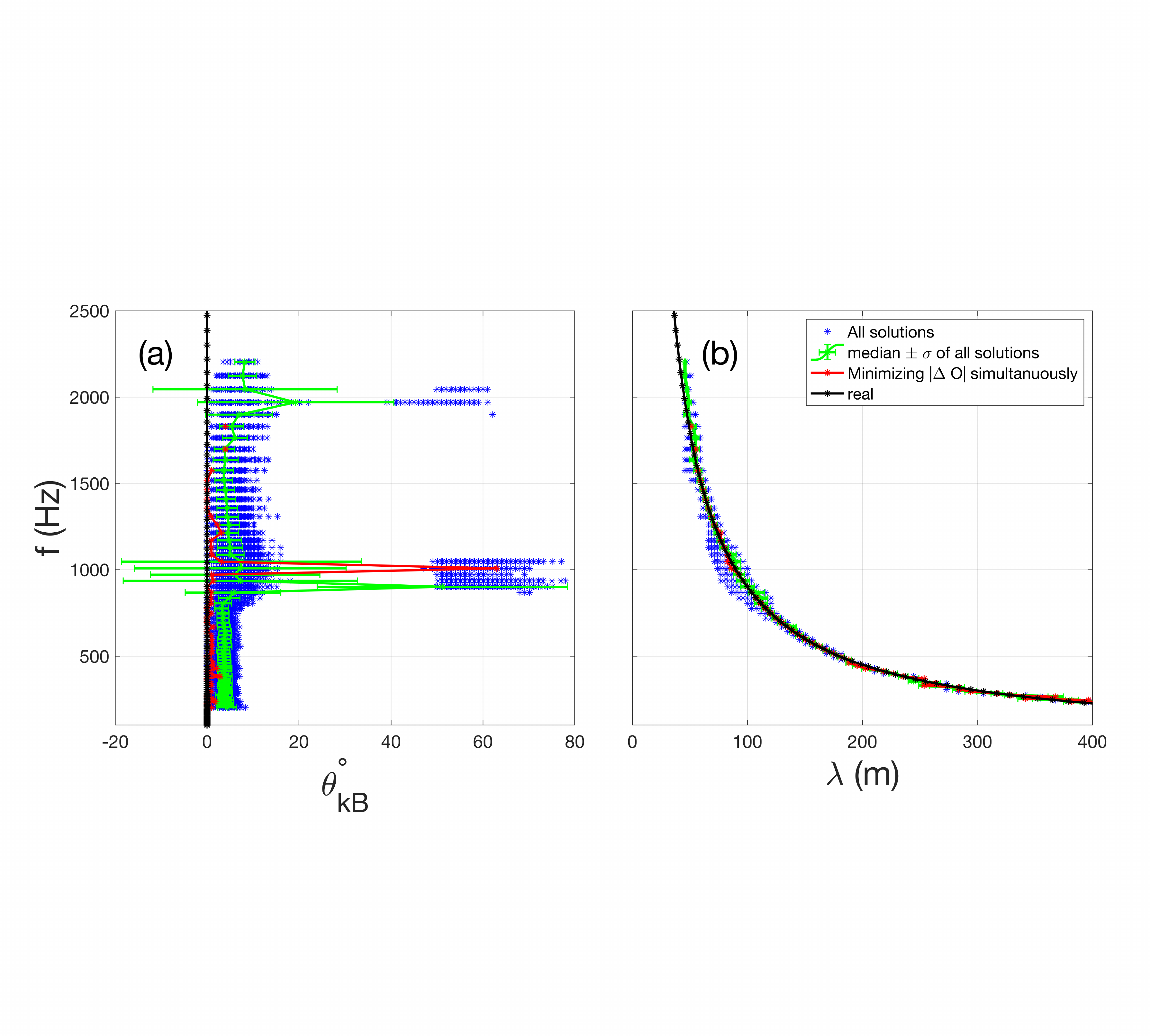}
    \caption{Solutions to the inverse problem. Panels (a) and (b) shows the frequency versus $\theta_{kB}$ and frequency versus $\lambda$ respectively. Blue stars are all the waves that solve the set of 5 equations, black line represent the real values, red are the results of finding the one solution that minimizes the set of 5 equations simultaneously and green errorbar represent the median and standard deviation of all solutions.}
    \label{fig:fig7}
\end{figure*}

Multiple triplet combinations satisfy equation \ref{condition}. We apply this calculation to the simulated wavepacket whose potential is described by equation \ref{eq4} with white noise with signal to noise ratio of 15 dB added to simulate real waves measurement. Figure \ref{fig:fig7} (a) shows f versus $\theta_{kB}$, while panel (b) is a plot of the f versus $\lambda$. In both panels the blue stars are the values corresponding to all solutions of the system of equations \ref{condition} at each frequency component, the black line is the expected values of $\theta_{kB}$/$\lambda$, the red line is the values that minimize all 5 equations simultaneously, and the green errorbars represent the median and standard deviation of all solutions at each frequency. The solution that minimizes all quantities simultaneously (red curves) does a good job at retrieving the dispersion relation, but at some points it deviates from the true properties of the wave (it predicts a $\theta_{kB} \sim 60^\circ$ at $f \sim 1$ KHz). So instead the final output of the method and what we report as our final measurement for the wave properties will be the median of all the solutions at each frequency component and we take the standard deviation of all solutions as an error estimate on this measurement (green errorbar). As is clear from panels (c-d) of Figure \ref{fig:fig6}, below a wavelength of  $\sim 50$ m the integer fraction of the wavelength where the method fails becomes closer to each others. Also, from Figure \ref{fig:fig2} it is clear that the assumption that the axial component of the electric field is not subject to attenuation fails. That is why we limit the simulation to wavelength larger than 45 m and take that value as the minimum wavelength that we can resolve.

\begin{figure*}[ht]
    \centering
    \includegraphics[scale=0.25]{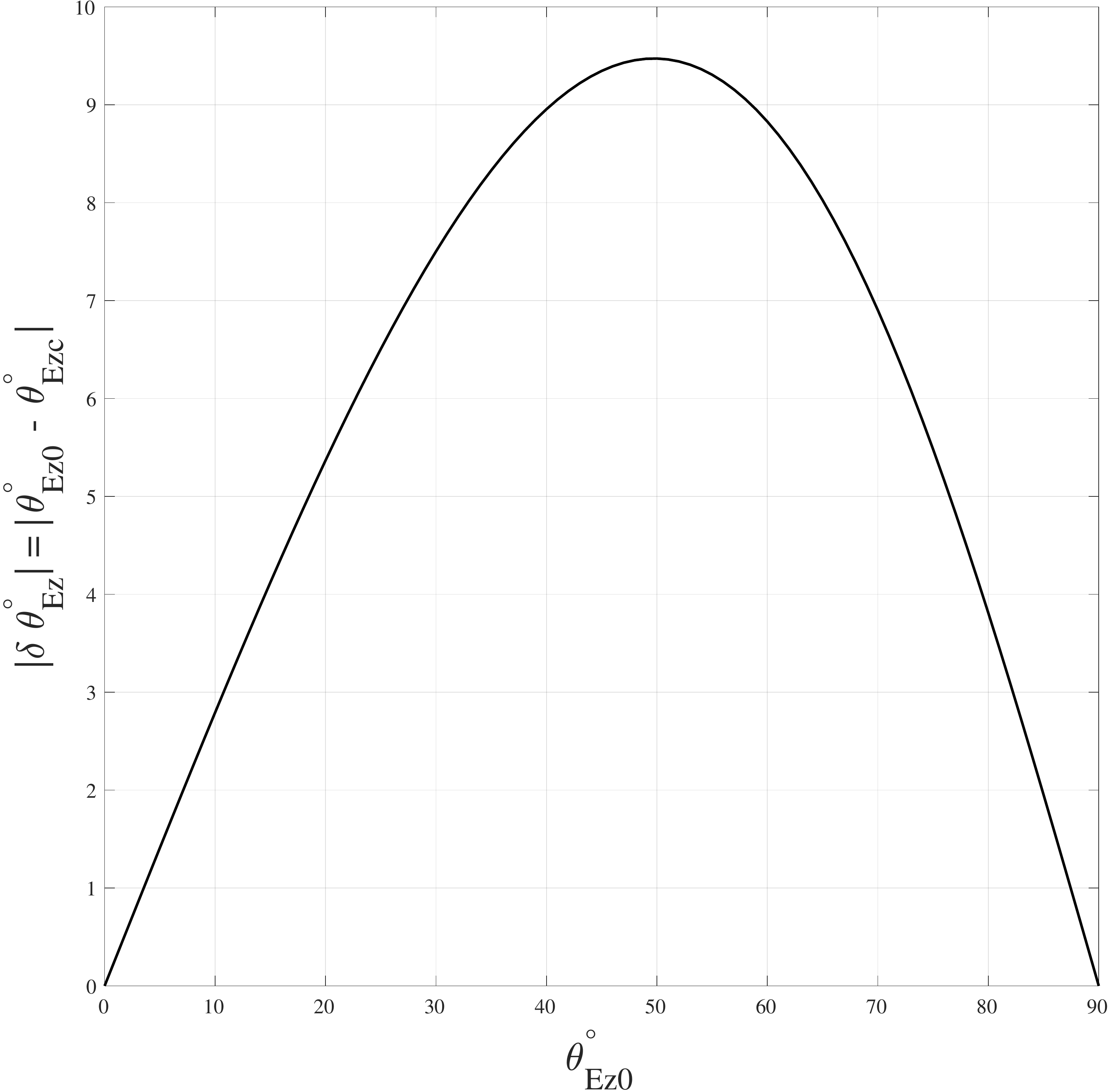}
    \caption{$|\delta \theta_{Ez} |$, uncertainty due to ignoring boom shorting and sheath impedance effects, versus polar angle $\theta_{Ez0}$.}
    \label{fig:fig8}
\end{figure*}
In developing this method we neglected any corrections in amplitudes due to sheath impedance or boom shorting effects. The usual way to correct the electric field is by multiplying its amplitude by gain factors corresponding to each effect. Those gain factors are dependent on density and frequency but usually they are assumed to be constant. For the boom shorting effects the value for the gain factors usually used are $2.1$ for the axial probes and $1.25$ for the spin plane probes \cite<see table 3 in >[]{ergun2016axial}. While for the sheath impedance effect multiple values have been reported for the gain factors, the most recent estimate was by \citeA{wang2021electrostatic} obtaining values of $1.6$ for the axial probes and $1.8$ for the spin plane probes. The difference between the axial and spin plane gain factors affect the value of $\theta_{Ez}$ that we use to get the 3D wave vector in our method. Using the gain factors listed above we can estimate what kind of error neglecting those effects introduce to our measurement. For unit vectors at a varying polar angles $\theta_{Ez0}$ and fixed azimuthal angle $\phi = 45^\circ$ we multiply the components by the relevant gain factors and renormalize the vector then calculate the polar angle of the vector after correction $\theta_{Ezc}$. We then calculate $|\delta \theta_{Ez} | = |\theta_{Ez0} - \theta_{Ezc}|$, and plot it as a function of $\theta_{Ez0}$ in Figure \ref{fig:fig8}. It is clear that the maximum variance in $\theta_{Ez}$ that not accounting for boom shorting and sheath impedance effects would add is around $9^\circ$, which can be considered as an additional uncertainty on our measurement.

\subsection{3D wave vector determination: Spacecraft data}
\begin{figure*}[ht]
    \centering
    \includegraphics[scale=0.3]{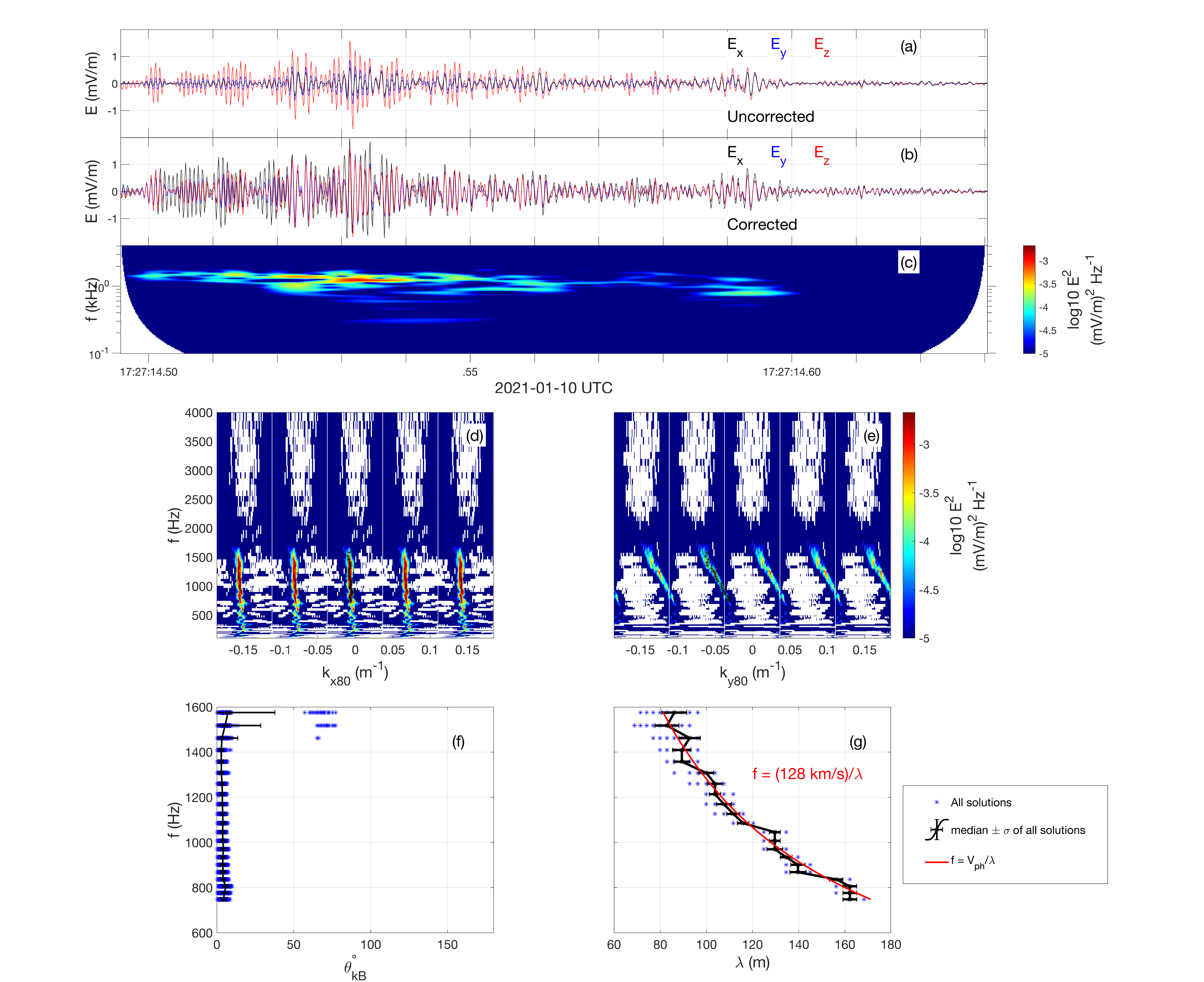}
    \caption{Application of the method to the waveburst in Figure \ref{fig:fig1}. Panel (a) shows the uncorrected electric field in the PCS, (b) corrected electric field, (c) uncorrected electric field PSD, (d - e) f - k PSD in the $x_{80}$ and $y_{80}$ directions respectively. Panels (f - g) shows the result of the method, where frequency versus $\theta_{kB}$ is plotted in (f) and versus $\lambda$ (g). Blue stars are all solutions matching the observables, black errorbar represent the median and standard deviation of those solutions. The red curve in panel (g) represent a weighted fit to the equation $f = V_{ph}/\lambda$, of the data with the standard deviation as the weights.}
    \label{fig:fig9}
\end{figure*}

Now that we developed the method on synthetic data, we apply it to the waveburst measured by MMS1 shown in Figure 1. In panel (a) of Figure \ref{fig:fig9} we plot the uncorrected electric field  showing that the amplitude of the z component of the electric field is significantly larger than that in the spin plane (x,y). Panel (b) shows electric field after correcting for the attenuation, using the method described above, in the spin plane. As is clear, the spin plane components of the electric field have now comparable amplitude to that in the axial direction. Panel (c) shows the uncorrected electric field PSD. Panels (d) and (e) shows the f - k PSD in the $x_{80}$ and $y_{80}$ directions showing a linear dispersion relation. As before, we choose the branches that connect to the origin and highlight them with the black stars.

\begin{figure*}[ht]
    \centering
    \includegraphics[scale=0.3]{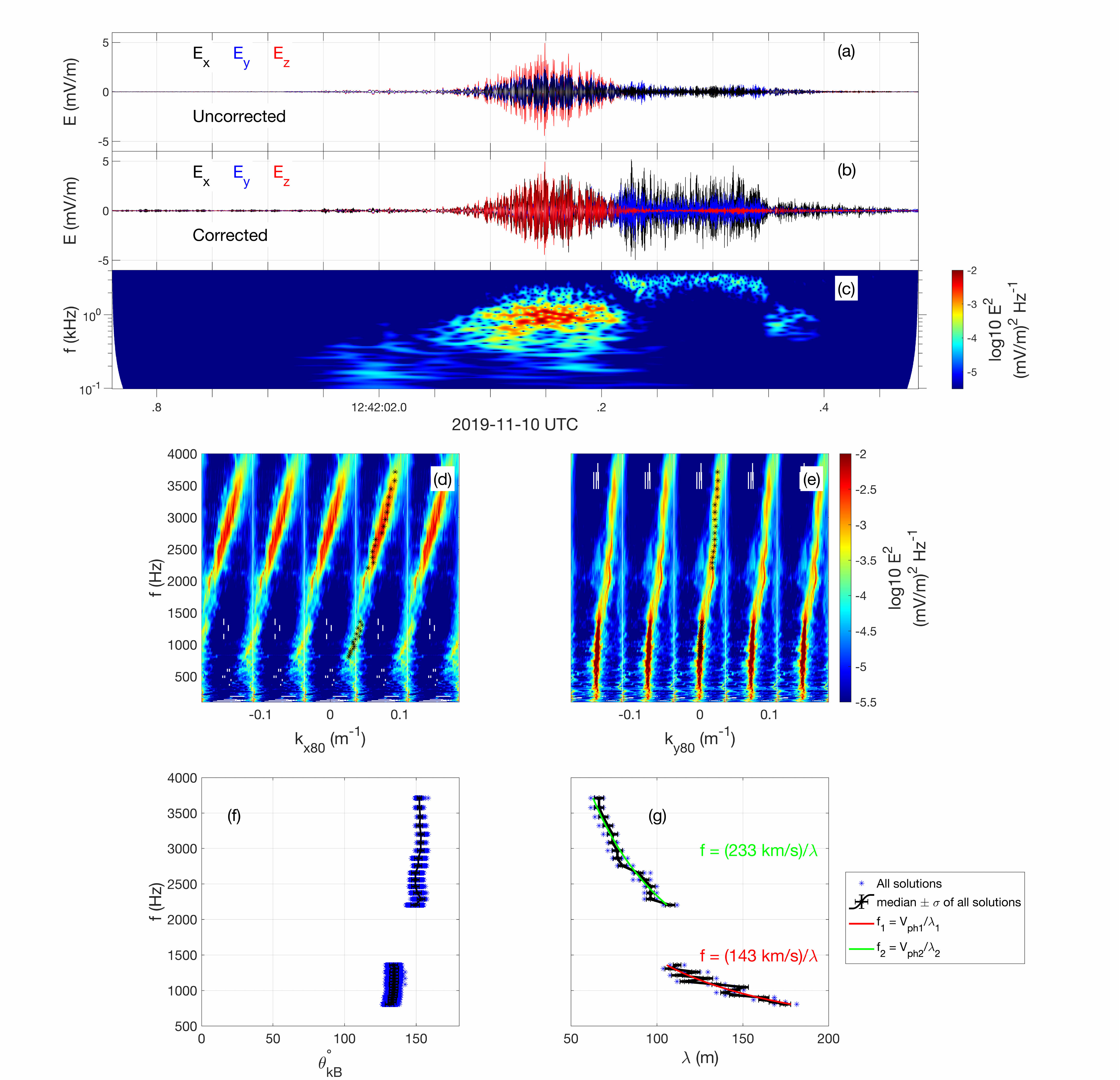}
    \caption{Waveburst showing a discontinuity in its PSD along with the measured dispersion relation. Same format as in Figure \ref{fig:fig9}.}
    \label{fig:fig10}
\end{figure*}
 Panel (f) shows the frequency versus $\theta_{kB}$, while panel (g) shows the frequency versus the wavelength, with least square fits to the function $f = V_{ph}/\lambda$ overlayed in red. On top of the measurement of the dispersion relation in panel (g), we clearly see in panel (f) that this waveburst is not oblique and travels in the field-aligned direction with $\theta_{kB} \sim 3^\circ$. This can be compared to $\theta_{kB} \sim 31^\circ$ obtained using the maximum variance direction of the uncorrected electric field. It is worth noting that at higher frequencies two different clusters of solutions are visible in panel (f), the first have $\theta_{kB} \sim 3^\circ$ and the second around $70^\circ$. The error bars for those frequencies are significantly larger compared to the other frequencies, as they should be, reflecting the extra uncertainty that the second cluster of solutions add to our measurement.

\section{Properties of ion acoustic waves in the solar wind}
\label{test}

In this section we use the method developed above to investigate the properties of the solar wind ion acoustic wavebursts. First, we look at an event where the waveburst exhibits interesting behavior in its wavelet spectrum. Second, we perform a statistical study of the properties of ion acoustic waves in the solar wind.

\subsection{Case study}

Several wavebursts in our list exhibit an interesting behavior in their PSD where what seems to be a continuous waveform in the time domain, exhibits a discontinuity in the wavelet domain. An example is shown in Figure \ref{fig:fig10}. Panel (a) shows the uncorrected electric field featuring an increasing amplitude peaking at around 5 mV/m then decreasing with what looks like a wavetrail. Panel (b) shows the corrected electric field showing that the lower amplitude wavetrail is nothing but part of the higher frequency component of the waveburst which is subject to great attenuation in amplitude. Panel (c) shows the uncorrected electric field power spectral density (PSD) which shows that what seems to be one waveburst has two disconnected PSD signatures one at lower frequency and the other at higher frequency. Panels (d) and (e) shows the f - k PSD in the $x_{80}$ and $y_{80}$ directions again showing the signature of two separate dispersion relations.

 Panels (f-g) shows the wave properties obtained for this waveburst. Panel (f) shows the frequency versus $\theta_{kB}$, while panel (g) shows the frequency versus the wavelength, with least square fits to the function $f = V_{ph}/\lambda$ overlayed. The green fit is for the higher frequency component of the dispersion relation while the red fit is for the lower frequency component. We can clearly see that the lower frequency component of the waveburst travel at a more oblique angle ($\theta_{kB} \approx 130^\circ$) and with slower phase velocity ($V_{ph} = 143$ km/s) in the spacecraft frame, compared to the higher frequency component travel which is more anti-field aligned ($\theta_{kB} \approx 150^\circ$) and at higher phase velocity ($V_{ph} = 233$ km/s). If we Doppler shift the dispersion relation to the plasma frame (not shown) we see that the two disconnected components become one continuous dispersion relation. This means that due to the different direction of propagation the two wave bursts appear discontinuous in the spacecraft frame. 


\subsection{Statistical study}

 From the previously compiled 210 wavebursts, only 105 show a clear dispersive character in the spin plane allowing us to apply the method. For each of those events we apply the method and fit the resulting dispersion relation to the equation $f = V_{ph}/\lambda$. To insure that the data properly fit the dispertion relation we use the parameter R-Squared = $1-\frac{\sum_i \left(y_i - f_i\right)^2}{\sum_i \left(y_i - \bar{y}\right)^2}$ to assess the goodness of the fit. In the formula for R-Squared, $y_i$ is the $i^{th}$ measured value of the dependent variable $y$ corresponding to the $i^{th}$ measured value of the independent variable $x$ in the fit, $f_i$ is the value of the fit evaluated at $x_i$ and $\bar{y}$ is the mean value of $y_i$. The closer R-squared is to 1 the better the fit is. We discard all events whose fits have an R-Squared value less than 0.1 (i.e the measured dispersion relation does not fit the function $f = V_{ph}/\lambda$ properly), which leaves us with a total of 48 events. For each of those 48 events we find the frequency value which corresponds to the peak power. The interpolated $\lambda$ and $\theta_{kB}$ at that frequency are then taken as the representative values for each waveburst.
 
\begin{figure*}[ht]
    \centering
    \includegraphics[scale=0.3]{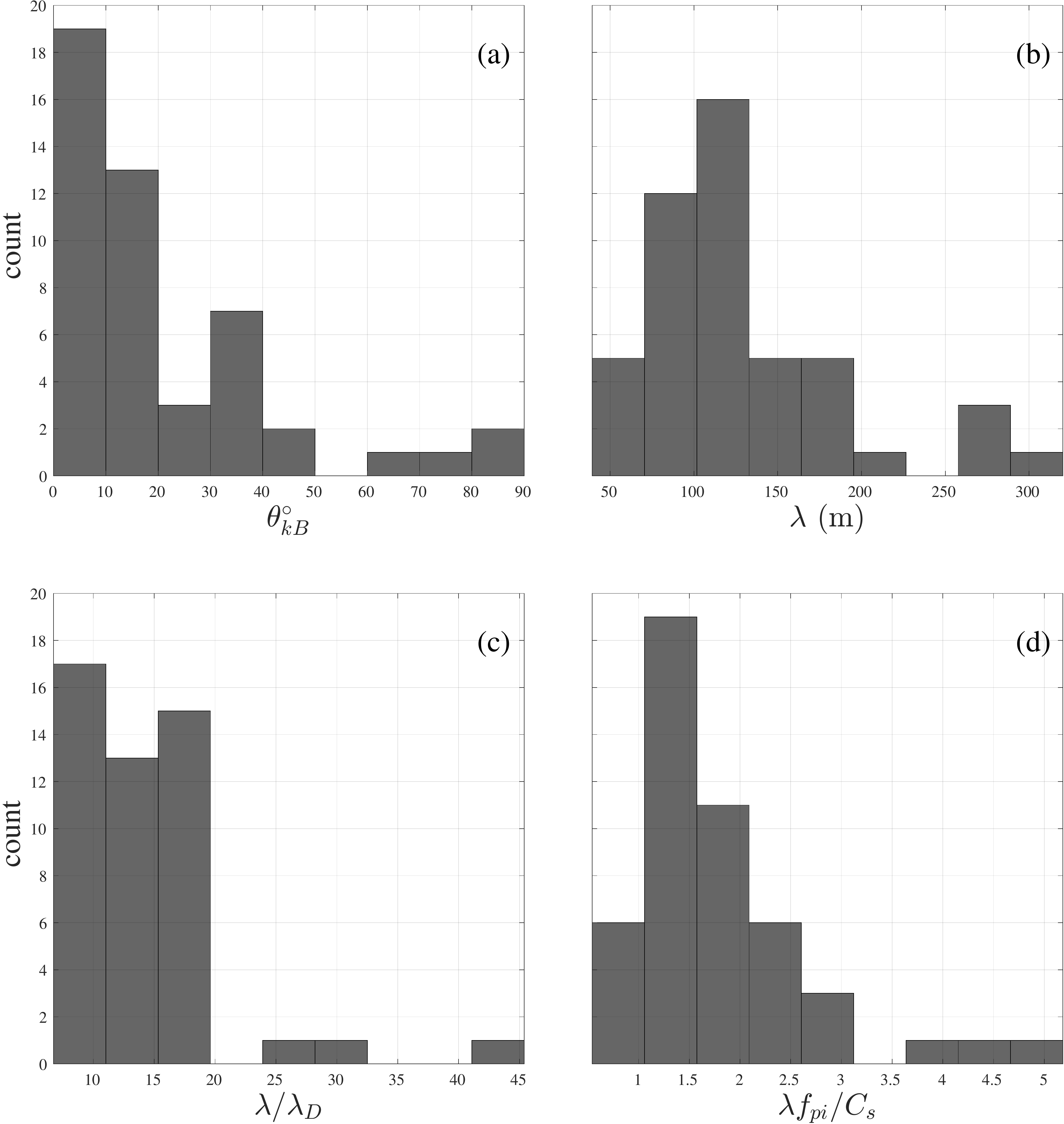}
    \caption{Statistical properties of ion acoustic waves in the solar wind. Panel (a) is a hystogram of $\theta_{kB}$, (b) histogram of the wavelength $\lambda$, (c) histogram of the wavelength normalized to the Debye length $\lambda/\lambda_D$ and (d) histogram of the wavelength normalized to $f_{pi}/C_s$ with $f_{pi}$ being the measured ion plasma frequency and $C_s$ the measured sound speed.}
    \label{fig:fig11}
\end{figure*}

 In Figure \ref{fig:fig11} we plot the histogram of the results. Panel (a) shows the distribution of $\theta_{kB}$. We retrieve the expected result that ion acoustic waves in the solar wind are predominantly field aligned. In panels (b) we show the statistical distribution of the wavelength of ion acoustic waves in the solar wind. The distribution peaks at around 100 m consistent with previous measurements \cite{1978Gurnettionacoustic}. PanelS (c - d) shows the distribution of wavelengths normalized to $\lambda_D$ and $f_{pi}/C_s$, respectively, where $f_{pi}$ is the measured ion plasma frequency and $C_s$ is the measured ion sound speed. The ratio $f_{pi}/C_s$ can be taken as a characteristic wavelength for ion acoustic waves. The measured distribution peaks between 10 and 20 $\lambda_D$ or between 1 and 2 $f_{pi}/C_s$, which is consistent with theoretical expectations.

\section{Conclusion}
\label{conclusions}

Characterization of high-frequency short-wavelength electrostatic waves requires reliable measurement of the wave electric field along with the determination of the wave vector. We observe that the electric field of ion-acoustic waves measured by MMS in the solar wind is systematically biased towards the axial (approximately GSE Z) direction. A similar problem has been identified for the high-frequency waves at the shock \cite{2018Goodrich}. This bias makes it difficult to determine the wave mode using the measured electric field.

We show that this bias is caused by the electric field measured by the double-probe instrument being attenuated when the wavelength of the waves approaches the probe-to-probe separation (short wavelength effect). To address this problem we develop a method to measure the 3D wave vector of an electrostatic wave. The method is based on spin-plane interferometry, it assumes that we are measuring plane waves with wavelength larger than 45 m (1.5 times separation of the axial probes, and 0.37 times that of the spin plane probes) and propagating at an angle to the axial direction so that there is a significant signal in the spin-plane measurement. We benchmark this method on both synthetic data and real data whose properties are generally known, namely ion acoustic waves measured in the solar wind. Previous statistical analysis of solar wind ion acoustic waves was done using 2D measurements of the electric field. Instead, our method allows us to determine the full 3 dimensional wave vector of the waves for the first time. We find that the waves travel predominantly in the field aligned direction (well known result), and have a wavelength of $\sim$100 m or 10 to 20 Debye lengths. 

The proposed method can be applied to study short-wavelength electrostatic waves (wavelength below $\sim$1000 m), which often occur in the near-Earth regions with sufficiently short Debye lengths encountered by MMS, particularly at the bow shock, magnetosheath and magnetopause.

\section{Open Research}
MMS data is available at \url{https://lasp.colorado.edu/mms/sdc/public/data/}. Data analysis was performed using the IRFU-Matlab analysis package. The code for applying the method developed in this paper along with the code to generate the simulation data can be found at \url{https://github.com/ahmadlalti/ESW-measurement.git}. The simulation data can be found at \url{https://doi.org/10.5281/zenodo.7310062}.
 \acknowledgments
We thank the entire MMS team and instrument PIs for data access and support. This work is supported by the Swedish Research Council grant 2018-05514.

\bibliography{biblio}


%
%




%
%
%
%
%

\end{document}